\documentclass[aps,prc,reprint,showpacs,preprintnumbers,amsmath,amssymb] {revtex4-1}
\usepackage{amsmath}
\usepackage{amssymb}
\usepackage{graphicx}
\usepackage{subfigure}
\usepackage{rotating}
\usepackage{natbib}
\usepackage{txfonts}
\usepackage{color}
\def\beq{\begin{equation}}
\def\eeq{\end{equation}}
\def\bea{\begin{eqnarray}}
\def\eea{\end{eqnarray}}

\def\eqref#1{Eq.~(\ref{eq:#1})}

\def\be{\begin{equation}}
\def\ee{\end{equation}}
\def\bg{\begin{eqnarray}}
\def\en{\end{eqnarray}}

\long\def\Omit#1{}

\DeclareGraphicsExtensions{ .pdf, .jpg, .eps}
\begin{document}
\title{Determination of the $^{36}$Mg($n,\gamma$)$^{37}$Mg reaction rate from Coulomb dissociation 
of $^{37}$Mg 
}
\author{Shubhchintak$^{1}$}
\email{Shub.Shubhchintak@tamuc.edu}
\author{R. Chatterjee$^2$}
\email{rcfphfph@iitr.ac.in}
\author{R. Shyam$^3$}
\email{radhey.shyam@saha.ac.in}
\affiliation{$^1$Department of Physics and Astronomy, Texas A$\&$M University-Commerce, Commerce, Texas 75428, USA}
\affiliation{$^2$Department of Physics, Indian Institute of Technology, 
Roorkee 247667, India}
\affiliation{$^3$Saha Institute of Nuclear Physics, 1/AF Bidhan Nagar, 
Kolkata 700064, India}

\date{\today}
\begin{abstract}
We use the Coulomb dissociation (CD) method to calculate the rate of the
$^{36}$Mg(${n,\gamma}$)$^{37}$Mg radiative capture reaction. The CD cross
sections of the $^{37}$Mg nucleus on a $^{208}$Pb target at the beam energy 
of 244 MeV/nucleon, for which new experimental data have recently become
available, were calculated within the framework of a finite range
distorted wave Born approximation theory that is extended to include the
projectile deformation effects. Invoking the principle of detailed 
balance, these cross sections are used to determine the excitation 
function and subsequently the rate of the $^{36}$Mg($n,\gamma$)$^{37}$Mg 
reaction. We compare these rates to those of the 
$^{36}$Mg($\alpha,n$)$^{39}$Si reaction calculated within a Hauser-Feshbach 
model.  We find that for $T_9$ as large as up to 1.0 (in units of 
10$^9$ K) the $^{36}$Mg($n,\gamma$)$^{37}$Mg reaction is much faster than 
the $^{36}$Mg($\alpha,n$)$^{39}$Si one. The inclusion of the effects of 
$^{37}$Mg projectile deformation in the breakup calculations, enhances 
the ($n,\gamma$) reaction rate even further. Therefore, it is highly unlikely
that the $(n,\gamma)$ $\beta$-decay $r$-process flow will be broken at the 
$^{36}$Mg isotope by the $\alpha$-process. 

\end{abstract} 
\maketitle

\section{Introduction}
It is generally believed that the $r$-process that synthesizes the heavy 
isotopes occurs under explosive conditions at high neutron densities 
and high temperatures~\cite{cow91,lan01,arn07}. The post-collapse phase 
of a type II or type Ib supernova can provide such a situation   
\cite{mey92,woo94}. In the early expanding phase, starting with a 
He-rich environment, the mass-8 gap would be bridged by either 
$\alpha + \alpha + \alpha \to ^{12}$C or $\alpha + \alpha + n \to ^{9}$Be 
reactions. The $\alpha$-capture reactions would continue until 
temperatures and densities become relatively low and the charged particle 
reactions almost cease. At this stage a very neutron-rich freeze-out 
takes place, which triggers further synthesis of the elements by the 
radiative neutron-capture process~\cite{ter01}. 

For calculations of the $r$-process nucleosynthesis the inclusion of 
neutron-rich light nuclei in the reaction network, has been shown to be 
important - they can change the heavy element abundances by as much as
an order of magnitude~\cite{ter01,tak05}. The $r$-process path involving 
neutron-rich nuclei can in principle, go up to the drip line isotope once 
equilibrium between $(n,\gamma)$ and $(\gamma,n)$ processes is established. 
If, however,  the $(\alpha,n)$ reaction becomes faster than the $(n,\gamma)$ 
reaction on some ``pre-drip line" neutron-rich isotope, then the $r$-process 
flow of the radiative neutron capture reaction followed by the $\beta$-decay 
is broken and the reaction path will skip the isotope on the drip line.  

The abundance yields of extremely neutron rich nuclei show that the largest 
abundance is exhibited by the isotope of a given atomic number $Z$ that is on 
or very close to the corresponding neutron-drip line. However, $^{18}$C and 
$^{36}$Mg are exceptions to this observation. Both these isotopes are still 
away from the respective drip lines of the corresponding $Z$ values. For the 
carbon nucleus the drip line isotope is known to be $^{22}$C~\cite{mos13,tos16}, 
while for magnesium the drip line is extended upto $^{40}$Mg~\cite{bau07}. It 
was speculated in Ref.~\cite{ter01} that even at low temperatures 
(around $T_9$ = 0.62), the rates of $^{18}$C($\alpha,n$)$^{21}$O and 
$^{36}$Mg($\alpha,n$)$^{39}$Si reactions can be larger than those of 
$^{18}$C($n,\gamma$)$^{19}$C and $^{36}$Mg($n,\gamma$)$^{37}$Mg reactions, 
respectively. To confirm this observation the precise determination of the 
rates of these reactions is of crucial importance.
 
The aim of this paper is to determine the rates of the 
$^{36}$Mg(${n,\gamma}$)$^{37}$Mg reaction at the interaction kinetic energies 
that correspond to temperatures in the astrophysically interesting region 
($T_9$ = 0.5 - 10). Since, $^{36}$Mg has a very small half-life ($\approx$ 
5 ms)~\cite{doo13}, a direct measurement of the cross section of the reaction 
$^{36}$Mg($n,\gamma$)$^{37}$Mg is not feasible at present. The situation is 
further complicated by the fact that temperatures $T_9$ with values in the 
region of 0.5 - 10 correspond roughly to the center-of-mass (c.m.) energies 
in the range of 50 keV to 1.0 MeV. Thus rates of the
$^{36}$Mg(${n,\gamma}$)$^{37}$Mg reaction are of astrophysical importance for 
such low neutron kinetic energies, where performing measurements is 
prohibitively difficult.

However, with a beam of $^{37}$Mg it is possible to measure the cross 
section of the reverse reaction $^{37}$Mg($\gamma,n$)$^{36}$Mg 
(photodisintegration process), and use the principle of detailed 
balance to deduce from it the cross sections of the 
$^{36}$Mg(${n,\gamma}$)$^{37}$Mg reaction. A very promising way of studying 
the photodisintegration process is provided by the virtual photons acting 
on a fast charged nuclear projectile passing through the Coulomb field of 
a heavy target nucleus~\cite{bau86,bau03,bau08}. The advantage of this 
Coulomb dissociation (CD) method, in which the valence neutron is removed 
from a fast projectile in the Coulomb field of heavy target nuclei, is that 
here measurements can be performed at higher beam energies, which enhances 
the cross sections considerably. At higher energies the fragments in the 
final channel emerge with larger velocities that facilitates their more 
accurate detection. Furthermore, the choice of the adequate kinematical 
condition of the coincidence measurements allows the study of the low 
relative energies of the final state fragments and ensures that the target 
nucleus remains in the ground state during the reaction (see, e.g., Refs.
\cite{sch06,nak99}). 

With the advent of new generation of radioactive ion beam facilities, it 
has become possible to produce a beam of $^{37}$Mg of sufficient quality to 
perform the measurements for the cross sections of the one-neutron removal
reaction off this nucleus on a $^{208}$Pb target at a beam energy of 244 
MeV/nucleon~\cite{kob14}. The corresponding data were analyzed within 
a post-form finite range distorted-wave Born approximation (FRDWBA) theory 
that is extended to include projectile deformation effects~\cite{shu14,shu15}. 
From a comparison of calculations with the available experimental data it was 
concluded that the likely configuration of the $^{37}$Mg ground state is either 
$^{36}$Mg(0$^+$)$\otimes$2$p_{3/2}n$ or $^{36}$Mg(0$^+$)$\otimes$2$s_{1/2}n$
with the one neutron separation energy $(S_n)$ values of 0.35 $\pm$ 0.06 MeV 
and 0.50 $\pm$ 0.07 MeV, respectively. These values were found to be 
strongly dependent on the spectroscopic factors ($C^2S$) and the deformation 
of the respective configuration.
  
In this work we use the Coulomb breakup cross section of $^{37}$Mg on a 
$^{208}$Pb target calculated within the FRDWBA theory as described above, to 
determine the photoabsorption cross sections $^{37}$Mg($\gamma, n$)$^{36}$Mg by 
following the method of virtual photon number~\cite{bau86}. The later was then 
converted to $(n,\gamma)$ capture cross section on $^{36}$Mg using the 
principle of detailed balance.

We adopt the cross sections of the $^{36}$Mg($\alpha,n$)$^{39}$Si reaction 
as obtained from the Hauser-Feshbach (HF) code NON-SMOKER in Ref.~\cite{rau01}.  

In the next section we present our formalism, where the features of the FRDWBA 
theory as used in the calculations of the Coulomb dissociation cross sections are 
briefly described. We also outline the virtual photon number method for extracting 
the photo-dissociation cross sections from those of the Coulomb dissociation
process. In Sec. III we present our results and discuss them. The summary and 
conclusions of our study are given in Sec. IV.  

\section{Formalism}

For non-degenerate stellar matter the rate ($R$) of a nuclear reaction where 
two nuclei $b$ and $c$ ($^{36}$Mg and $n$, respectively, in our example) 
interact with relative energy $E_{bc}$ to form a composite nucleus $a$ 
($^{37}$Mg) via a radiative capture process (to be represented as 
[$b (c,\gamma) a$], is given by (see, e.g., Refs.~\cite{rol88,tak05}),
\begin{eqnarray}
R &=& N_A\langle \sigma_{c\gamma}(v_{bc})v_{bc}\rangle, \label{r1} 
\label{Eq.1}
\end{eqnarray}
where $\sigma_{c\gamma}(v_{bc})$ is the cross section of the reaction between 
nuclei $b$ and $c$ with the relative velocity $v_{bc}$ (which corresponds to 
relative energy $E_{bc}$), and $N_A$ is the Avogadro number (= $6.02\times 
10^{23}$ mole$^{-1}$). In Eq.~(\ref{r1}) the product $\sigma_{c\gamma}(v_{bc}) 
v_{bc}$ is averaged over the Maxwell- Boltzmann velocity distribution.  
$\langle \sigma_{c\gamma}(v_{bc})v_{bc}\rangle$ is written as 
\begin{eqnarray}
\langle \sigma_{c\gamma}(v_{bc}) v_{bc} \rangle &=& \sqrt{\frac{8}{(k_B T)^3 \pi\mu_{bc}}}\int^{\infty}_0 \sigma_{c\gamma}
(E_{bc})\,E_{bc}\nonumber\\
&\times& exp(-\frac{E_{bc}}{k_BT})\,dE_{bc}, \label{Eq.2}
\end{eqnarray}
where $\mu_{bc}$  is the reduced mass of the interacting nuclei, $k_B$ is the 
Boltzmann constant, and $T$ is the temperature of the stellar medium. 

The prime nuclear physics input for calculating the rate of a particular 
radiative capture reaction is the cross section $\sigma_{c\gamma}(E_{bc})$  
at the relative energy $E_{bc}$ - this energy is usually in the range of 
a few keV to a few MeV for most of the astrophysical sites. The direct 
measurement of this cross section in the laboratory is extremely difficult at 
these low energies. Even more serious is the fact that for most reactions of 
interest the target nuclei are radioactive, having very short half lives.
 
However, by invoking the principle of detailed balance, the capture cross 
section ($\sigma_{c\gamma}$), can  be calculated from the cross section  
$\sigma_{\gamma c}$ of the time-reversed reaction $ a(\gamma,c)b$, 
(photodisintegration cross section of $a$) as,  
\begin{eqnarray}
\sigma_{c\gamma} = \frac{2(2j_a+1)}{(2j_b+1)(2j_c+1)}\frac{k_{\gamma}^2}
{k^2_{bc}} \sigma_{\gamma c}\label{Eq.3}, 
\end{eqnarray}
where $j_a$, $j_b$, and $j_c$ are the spins of  particles $a$, $b$, and $c$, 
respectively. $k_\gamma$ is the photon wave number and $k_{bc}$ is that of 
the relative motion between $b$ and $c$.

Now, the two-body photodisintegration cross section can be related to the relative energy spectra of the three-body elastic Coulomb breakup reaction 
($a+t \rightarrow b+c+t$, $t$ being a heavy target) as (see, e.g., 
Ref.~\cite{pra08})
\begin{eqnarray}
\frac{d\sigma}{dE_{bc}} = \frac{1}{E_\gamma}  \Sigma_{\lambda} n_{\Pi \lambda} 
\sigma_{\gamma c}, \label{Eq.4}
\end{eqnarray}
where $n_{\Pi \lambda}$ is the virtual photon number of type $\Pi$ (electric 
or magnetic) and multipolarity $\lambda$. The photon energy is given by 
$E_{\gamma} = E_{bc} + S_n$, with $S_n$ being the nucleon separation energy of 
the projectile $a$. The relative energy between the fragments $b$ and $c$ in 
the final channel is denoted by $E_{bc}$. 

Combining Eqs. (3) and (4) we can express $\sigma_{c\gamma}$ in terms of 
$\frac{d\sigma}{dE_{bc}}$ as

\begin{eqnarray}
\sigma_{c\gamma} = \frac{2(2j_a+1)}{(2j_b+1)(2j_c+1)} 2\mu_{bc} 
\frac{E_\gamma^3}{E_{bc}}\,\frac{1}{n_{\Pi \lambda}}\,\frac{d\sigma}{dE_{bc}},
\label{Eq.5}  
\end{eqnarray}
where we assume that the Coulomb breakup cross section gets a contribution
from a single multipolarity and type, $\Pi \lambda$. 

For application to the calculations of the reaction of interest in the present 
work, we use a fully quantum mechanical theory of Coulomb breakup reactions to 
calculate the Coulomb dissociation of $^{37}$Mg, which is then used to 
extract the rate of the capture reaction $^{36}$Mg($n, \gamma$)$^{37}$Mg. The 
theory of CD reactions used by us is formulated within the post-form (FRDWBA)~\cite{cha00}, where the 
electromagnetic interaction between the fragments and the target nucleus is 
included to all orders and the breakup contributions from the entire 
non-resonant continuum corresponding to all the multipoles and the relative
orbital angular momenta between the fragments are taken into account. This 
theory was extended in Refs.~\cite{shu14,shu15} so that it can also be 
used to calculate the CD of those nuclei that have deformed ground states. 
Full ground state wave function of the projectile, of any orbital angular 
momentum configuration, enters as an input into this theory, where we 
explicitly require only the ground state wave function of the projectile 
as an input.

Within the FRDWBA theory the cross sections for relative energy spectra 
for the elastic breakup reaction, $a+t\rightarrow b+c+t$, where projectile 
$a$ (assumed to have a core $b$ plus a valence particle $c$ configuration) 
breaks up into fragments $b$ and $c$ in the Coulomb field of a target $t$, 
can be written as,
\begin{eqnarray}
\frac{d\sigma} {dE_{bc} }&=&\int_{\Omega _{bc},\Omega_{at}} d\Omega _{bc} 
d\Omega_{at} \left\{\sum_{l m}\frac{1}{(2\ell + 1)}\vert \beta_{\ell m}\vert^2 
\right\} \nonumber\\
&\times&\frac{2\pi}{\hbar v_{at}} \frac{{\mu_{bc}\mu_{at}p_{bc}p_{at}}} 
{{h^6}}\label{Eq.6},
\end{eqnarray}
where $v_{at}$ is the $a-t$ relative velocity in the entrance channel,
$\Omega _{bc}$ and $\Omega_{at}$ are solid angles, $\mu_{bc}$ and $\mu_{at}$
are reduced masses, and $p_{bc}$ and $p_{at}$ are appropriate linear momenta
corresponding to the $b-c$ and $a-t$ systems, respectively. $\ell$ and 
$m$ are the relative orbital angular momentum and its projection, respectively. 
It may be noted that the projectile $a$ can also be deformed.

If one of the fragments (say $c$) is uncharged, the reduced transition
amplitude, $\beta_{\ell m}$, for this reaction is given by \cite{shu15}
\begin{eqnarray}
\beta_{\ell m} &=& \left\langle e^{i(\gamma{\bf q_c} - \alpha{\bf K}).{\bf r_1}}
\left|V_{bc}({\bf{r}}_{1})\right|\phi_{a}^{\ell m}({\bf r}_1)\right\rangle \nonumber\\ 
&\times&\left\langle \chi_{b}^{(-)}({\bf q_b},{\bf r_i})e^{i\delta 
{\bf q_c}.{\bf r_i}}|\chi_{a}^{(+)}({\bf q}_a,{\bf r_i})\right\rangle\label{Eq.7}.
\end{eqnarray}

The ground state wave function of the projectile $\phi_{a}^{\ell m}({\bf r}_1)$ 
appears in the first term (vertex function), while the second term that describes 
the dynamics of the reaction, contains the Coulomb distorted waves $\chi^{(\pm)}$.
This can be expressed in terms of the bremsstrahlung integral. $\alpha$, $\gamma$,
and $\delta$ are the mass factors pertaining to the three-body Jacobi coordinate
system (see Fig.~1 of Ref.~\cite{shu15}). In Eq. (\ref{Eq.7}), ${\bf K}$ is an 
effective local momentum appropriate to the core-target relative system and 
${\bf q}_i$ ($i = a, b, c$) are the Jacobi wave vectors of the respective 
particles.

$V_{bc}({\bf{r}}_{1})$ [in Eq. (\ref{Eq.7})] is the interaction between $b$ and 
$c$, in the initial channel. We introduce an axially symmetric quadrupole-deformed
potential, as
\begin{eqnarray}
V_{bc}({\bf r}_1) = {V_{0}}f(r_1)  
 -\beta_2 RV_{0} \frac{df(r_1)}{dr_1} Y^{0}_{2}(\hat {\bf r}_1), \label{Eq.8}
\end{eqnarray}
where $V_{0}$ is the depth of the spherical Woods-Saxon potential and $f(r_1) = [1+exp(\frac{r_1-R}{a_{0}})]^{-1}$, with $R = r_0A^{1/3}$, $r_0$ and $a_{0}$, the radius and diffuseness parameters, respectively. 
$\beta_2$ is the quadrupole deformation parameter. The first part of Eq. 
(\ref{Eq.8}) is the spherical Woods-Saxon potential $V_s(r_1)$. Because of the 
deformation, the radial wave function of a given $\ell$ corresponding to the 
full potential $V_{bc}$, has an admixture of wave functions corresponding the 
other $\ell$ values of the same parity. However, we calculate the radial part 
of the ground state wave function of the projectile using the undeformed 
Woods-Saxon potential $-$ this allows us to evaluate the structure part of the 
amplitude in Eq. (\ref{Eq.7}) analytically. This approximation is justified 
because it has been shown in Ref.~\cite{ham04} that in a realistic deformed 
potential the relative motion wave function of the neutron is dominated by the 
lowest angular momentum component in the limit of small binding energy of the 
valence neutron, which is independent of the extent of the deformation. 
\begin{figure}[t]
\centering
\includegraphics[width=.45\textwidth]{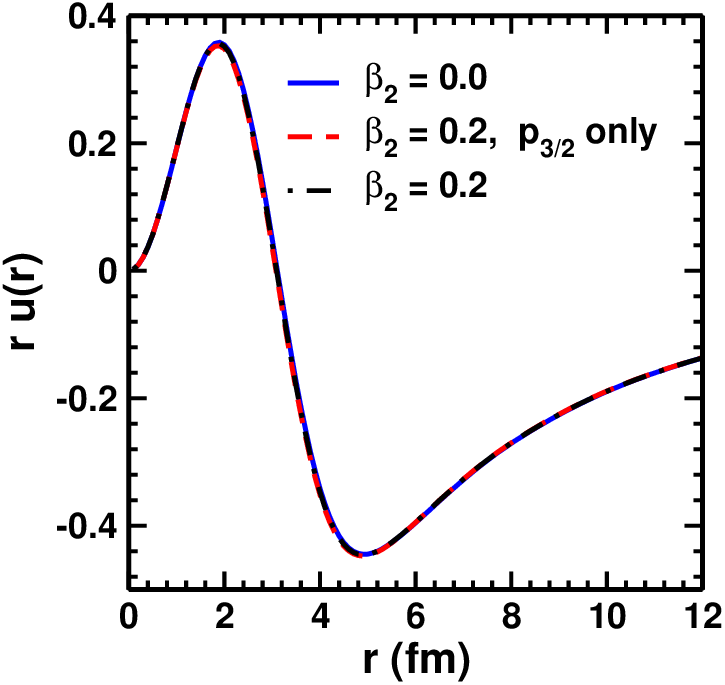}
\caption{
Comparison of the wave functions calculated by solving the Schr\"odinger 
equation with potential given by Eq.~(7). The $S_n$ value is taken to be 
0.35 MeV for the state $2p_{3/2}$. The wave function [$ru(r)$] obtained 
with the deformations parameter $\beta_2$ = 0.0 (i.e., calculated with 
spherical potential) is shown by the solid line. $ru(r)$ obtained 
with $\beta_2$ = 0.2 but including only the component corresponding to $\ell$ 
= 1 and $j$ = 3/2 is shown by the dashed line while that including components 
with $\ell$ = 1, 3, 5 and all the allowed $j$ values is displayed by the 
dashed-dotted line. All the wave functions are normalized to unity.
}
\label{fig:Fig1}
\end{figure}

To substantiate this point further, we show in Fig.~1 the wave function [ru(r)]
obtained by solving the Schr\"odinger equation with the potential given by Eq.~(8),
taking the $S_n$ value of 0.35 MeV. All the parameters of the potential were 
taken to be the same as those described above. We show results for $\beta_2$ = 
0.0 (solid line) and 0.2 (dashed and dashed-doted lines). The dashed line 
corresponds to the case when for $\beta_2$ = 0.2, only $\ell = 1$ and 
$j = 3/2$ component was included, while the dashed-dotted line represents the 
wave function where for $\beta_2$ = 0.2, all the components corresponding to 
$\ell = 1, 3$, and $5$ with all the allowed $j$ values are included. All the three
wave functions are normalized to 1 to make the comparison easier. We see that 
the solid and dashed curves are almost identical. The differences between the solid 
and dashed-dotted curves are also insignificant. This signifies that the 
deformation effects leave the wave functions calculated with the spherical 
potentials unchanged. Furthermore, contributions of $\ell > 1$ components to
$2p_{3/2}$ wave function are negligibly small.   

Therefore, we perform our CD calculations with the spherical wave function 
corresponding to the orbital angular momentum of 1 ($2p_{3/2}$ component), but 
taking Eq.~(8) for the potential $V_{bc}$. One advantage of our this choice is 
that this allows us to calculate a substantial portion of our amplitude 
analytically. We emphasize that the deformation parameter ($\beta_2$) still 
enters into the amplitude via $V_{bc}$ in Eq. (\ref{Eq.7}). For more details on 
the Coulomb dissociation formalism we refer to Ref.~\cite{shu15}.

One can then relate the cross section in Eq. (\ref{Eq.6}) to the photodissociation 
cross section, $\sigma_{\gamma c}$, for the reaction $a (\gamma, c) b$, by using 
Eq. (\ref{Eq.4}). The virtual photon number appearing in this equation was 
calculated by following the same method as that used in Ref.~\cite{ber99}.  

Of course, the procedure of relating the CD cross section to that of the  
photodisintegration is valid only when transitions of a single multipolarity 
and type dominate the breakup cross section and the nuclear breakup effects are 
negligible. The validity of both these assumptions has been checked in several 
previous studies of the Coulomb dissociation reactions (see, eg., Refs.
\cite{bau03,mot98,shy99,pra08,ber14,nee15}). Nevertheless, in the context of the 
reaction studied in this paper, we checked that the breakup cross sections are 
indeed dominated by the $E1$ multipolarity by evaluating the Coulomb dissociation 
cross sections within the first order Coulomb excitation theory~\cite{ber99}. 
Because the higher order effects that are included in the FRDWBA theory are 
negligible at higher beam energies as is shown in Ref.~\cite{pra02}, this procedure 
should be sufficient to satisfy that the FRDWBA cross sections are indeed dominated 
by the $E1$ multipolarity. Furthermore, at higher beam energies and forward angles 
(where the breakup data studied in Ref.~\cite{shu15} were taken), the nuclear 
breakup effects are negligible. Therefore, necessary conditions for the validity of 
Eq. (\ref{Eq.5}) are fulfilled for the present case. We emphasize, however, that, in 
general, the validity of Eq.~(\ref{Eq.5}) must be checked in each case before using 
this to extract the photodissociation cross section.

Once the photodissociation cross sections are determined, one can extract the 
radiative capture cross sections by using Eq. (\ref{Eq.3}) and use them in 
Eq. (\ref{Eq.2}) to determine the rate of the reaction.  

\section{Results and Discussions}

As discussed above, in the calculations of the CD cross sections within our 
theory, we require the single-particle wave function that describes the $c-b$
relative motion in the ground state of the projectile for a given 
neutron-core configuration. This is obtained by solving the Schr\"odinger 
equation with a central Woods-Saxon type of potential with parameters $r_0$ 
and $a_{0}$ having values of 1.24 and 0.62 fm, respectively. The depth of this 
well is adjusted to reproduce the valence neutron separation energy corresponding 
to the adopted configuration. In Ref.~\cite{shu15}, it was concluded that the 
ground state of $^{37}$Mg can have either of the configurations $^{36}$Mg(0$^+$)
$\otimes$2$p_{3/2}n$ and $^{36}$Mg(0$^+$)$\otimes$2$s_{1/2}n$, with $S_n$ values 
of 0.35 $\pm$ 0.06 MeV or 0.50 $\pm$ 0.07 MeV, respectively and a $C^2S$ of 1.
\begin{figure}[t]
\centering
\includegraphics[width=.45\textwidth]{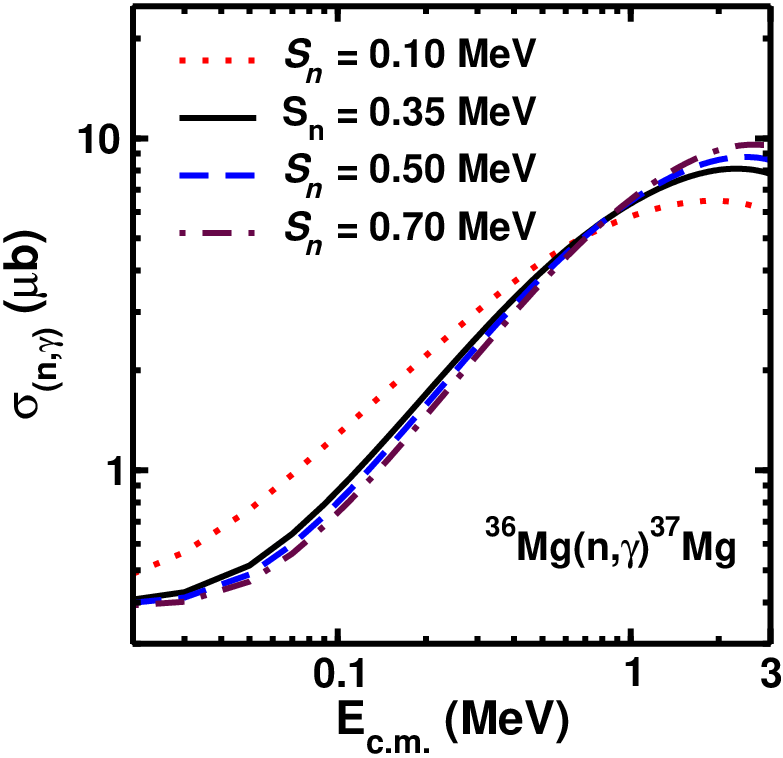}
\caption{
Direct capture (DC) cross sections to the ground state (GS) of $^{37}$Mg,
obtained by calculating the Coulomb dissociation of $^{37}$Mg on a
$^{208}$Pb target at the beam energy of 244 MeV/nucleon for different values
of one neutron separation energies ($S_n$). Results for $S_n$  values of 0.10, 0.35, 0.50 and 0.70 MeV, are shown by dotted, solid, dashed,
and dashed-dotted lines, respectively. In these calculations the deformation
parameter of the $^{37}$Mg ground state was taken to be 0 through out. The
spectroscopic factor $C^2S$ was unity in each case.
}
\label{fig:Fig2}
\end{figure}
\noindent 
However, for the configuration $^{36}$Mg(0$^+$)$\otimes$2$p_{3/2}n$ the calculated 
one-neutron removal cross section overlaps with the corresponding experimental data 
band for the quadrupole deformation parameter ($\beta_2$) below 0.32, which is in 
line with the predictions of the Nilsson model calculations of Ref.~\cite{ham07}. 
On the other hand, with the configuration $^{36}$Mg(0$^+$)$\otimes$2$s_{1/2}n$, 
the calculations of Ref.~\cite{shu15} is unable to put any constraint on the 
parameter $\beta_2$. Therefore, we adopt the  configuration $^{36}$Mg(0$^+$)
$\otimes$2$p_{3/2}n$ with a $S_{n}$ value of 0.35 MeV and a $C^2S$ of 1 for the 
$n-^{36}$Mg relative motion in the ground state of $^{37}$Mg. The 
values of the searched depths of the Woods-Saxon well with shape parameters 
as given above, were found to be 44.42, 45.21, 45.94, and 46.60 
MeV for $S_n$ values of 0.10, 0.35, 0.50, and 0.70 MeV, 
respectively. It should, however, be mentioned here that the extracted value 
of $S_n$ is sensitively dependent on $C^2S$ as well as on the deformation 
parameter $\beta_2$ of the $^{37}$Mg ground state. Since, definite knowledge 
about the later two quantities are still lacking we have chosen to show 
results for a range of $S_n$ and $\beta_2$ values.
\begin{figure}[t]
\centering
\includegraphics[width=.45\textwidth]{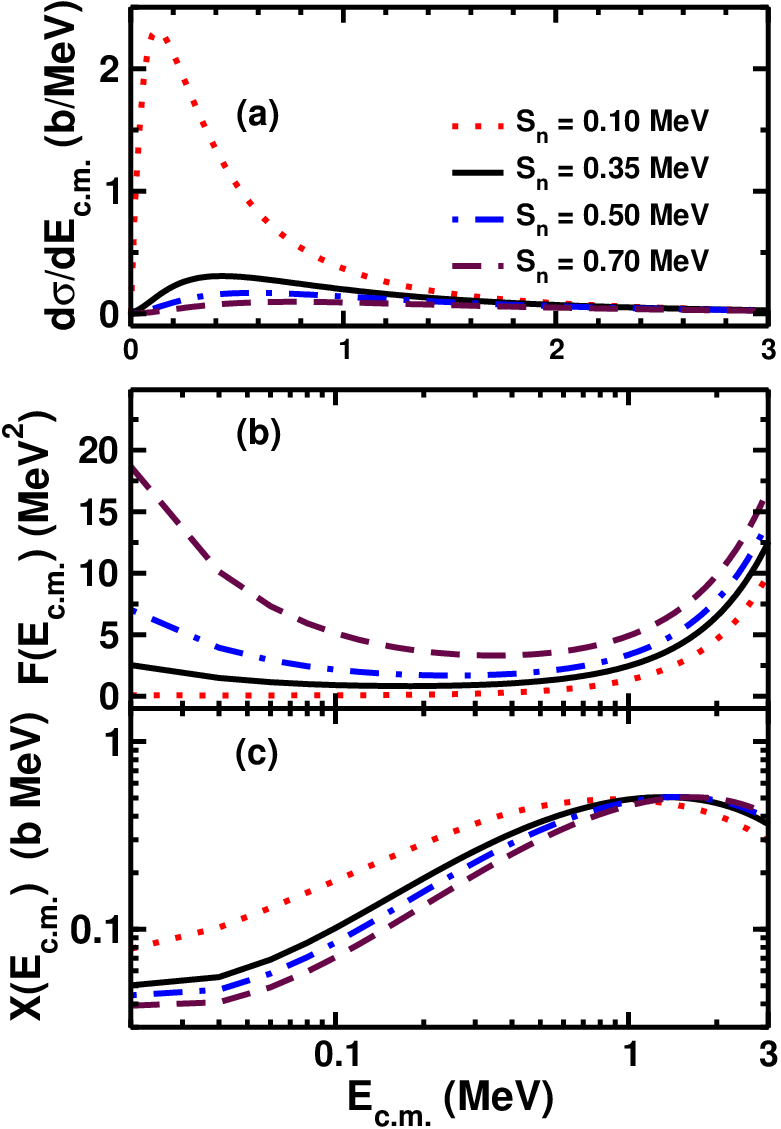}
\caption{
(a) The Coulomb dissociation cross section $d\sigma/dE_{c.m.}$ calculated for 
different values of $S_n$ as a function of $E_{\text{c.m.}}$. (b) The kinematical 
factor $F(E_{\text{c.m.}})$ = [$(E_{\text{c.m.}} + S_n)^3/E_{\text{c.m.}}$] as a function of 
$E_{\text{c.m.}}$ for various values of $S_n$. (c) The product of $F(E_{\text{c.m.}})$ and   
$d\sigma/dE_{\text{c.m.}}$ as a function of $E_{\text{c.m.}}$ for various values of $S_n$.  
}
\label{fig:Fig3}
\end{figure}

We calculated the capture cross sections ($\sigma_{n,\gamma}$) of the 
$^{36}$Mg($n, \gamma$)$^{37}$Mg reaction as a function of the c.m. relative 
energy ($E_{bc}$ = $E_{\text{c.m.}}$) between the neutron and $^{36}$Mg ground 
state [$^{36}$Mg($0^+$)] for several values of $S_n$ and $\beta_2$, using 
the Coulomb breakup cross section obtained in our FRDWBA model. In Fig.~2, 
we show $\sigma_{n,\gamma}$ as a function of $ E_{\text{c.m.}}$ (in the range 
of $0 - 3$ MeV) for $S_n$ values of 0.10, 0.35, 0.50, and 0.70 MeV 
corresponding to a fixed $\beta_2$ parameter of 0.0. We note in this figure
that while for $E_{\text{c.m.}}$ below 1 MeV, $\sigma_{n,\gamma}$ are larger 
for smaller values of $S_n$, this trend is reversed for $E_{\text{c.m.}}$
larger than 1 MeV, where the cross sections increase with increasing $S_n$.
The reason for this observation is discussed below. 

To understand the behavior of the $\sigma_{n,\gamma}$ as a function of 
${\rm E_{c.m.}}$ and $S_n$ as seen in Fig.~2, we note from Eq.~(5) that 
the capture cross sections obtained from the Coulomb dissociation method 
involves together with the CD cross section, the kinematical factor $F(E_{\text{c.m.}})$ 
= [$(E_{\text{c.m.}} + S_n)^3/E_{\text{c.m.}}$] and the inverse of the virtual photon number 
$n_{\Pi \lambda}$. In Fig.~3, we show the CD cross section [Fig. 3(a)], the 
kinematical factor $F(E_{\text{c.m.}})$ [Fig. 3(b)], and their product, $X(E_{\text{c.m.}})$ 
[Fig. 3(c)], as a function of $E_{\text{c.m.}}$ for various values of $S_n$. 

The Coulomb dissociation cross section shows the characteristics typical of 
the drip line nuclei having small one-neutron separation energies, where the 
breakup cross section is dominated by the low-lying dipole $B(E1)$ strength 
(see, e.g., Ref.~\cite{nag05}), which leads to these cross sections peaking 
strongly near the smaller binding energies. This implies that a low-lying 
bound state leads to a peak in the low lying continuum, and the width and 
location of that peak is directly related to the location of the bound state 
pole. As the binding energy changes, the strength distribution changes in 
both the shape and the absolute value, which is apparent from Fig. 3(a). 
\begin{figure}[t]
\centering
\includegraphics[width=.45\textwidth]{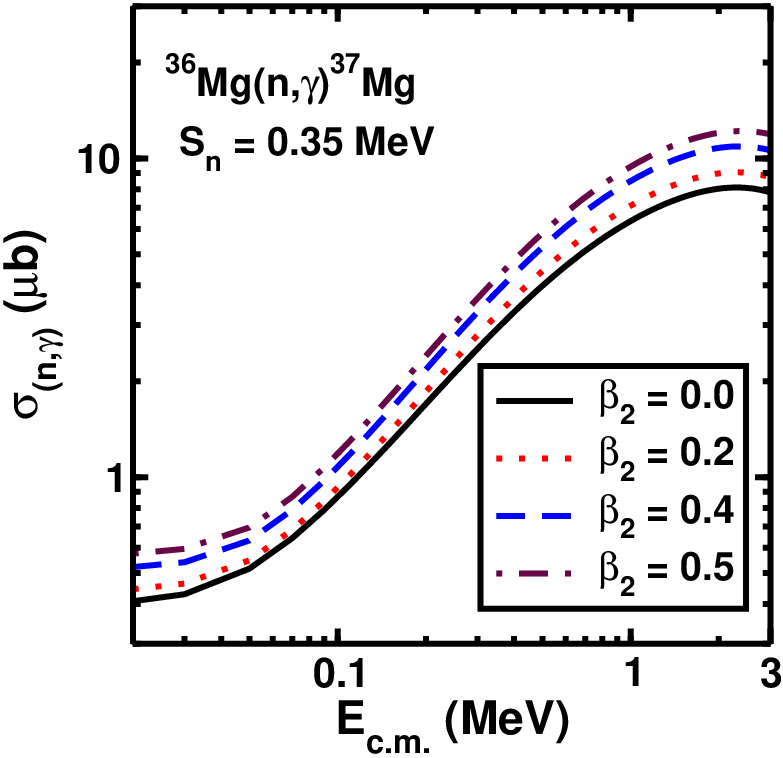}
\caption{
Same as in Fig.~2 obtained by using CD cross sections calculated with different 
$\beta_2$ values for a fixed $S_n$ of 0.35 MeV. Results for $\beta_2$  values of
0.0, 0.2, 0.4, and 0.5, are shown by solid, dotted, dashed, and dashed-dotted 
lines, respectively.  
}
\label{fig:Fig4}
\end{figure}

In Fig. 3(b), we have shown the kinematical factor $F(E_{\text{c.m.}})$  as a 
function of $E_{\text{c.m.}}$ for various values of $S_n$. We note that $F(E_{\text{c.m.}})$ 
is smallest in magnitude for the lowest value of $S_n$ and it increases 
gradually with $E_{\text{c.m.}}$ after some very small values of $E_{\text{c.m.}}$. As a result 
the product of $F(E_{\text{c.m.}})$ and the CD cross sections [$X(E_{\text{c.m.}})$], is still 
larger for smaller $S_n$ but only for $E_{\text{c.m.}} < 1$ MeV. However, for $E_{\text{c.m.}}$
larger than this value, this behavior is reversed - now $X(E_{\text{c.m.}})$ 
corresponding to the larger $S_n$ becomes larger. This is understandable because
the CD cross section remains approximately constant for $E_{\text{c.m.}} > 1$ MeV,
while $F(E_{\text{c.m.}})$ is bigger for larger values of $S_n$. Furthermore,
the virtual photon numbers have larger magnitudes for smaller $S_n$ for 
$E_{\text{c.m.}} < 1$ MeV, but they are of almost of similar values for all $S_n$ for 
$E_{\text{c.m.}} > 1$ MeV. This combined with $X(E_{\text{c.m.}})$ leads to the behavior
of the capture cross sections shown in Fig. 2, which appears to have a 
different $S_n$ dependence as compared to that of the Coulomb dissociation 
cross sections particularly for $E_{\text{c.m.}} > 1$ MeV. However, the capture cross 
sections corresponding to lower $S_n$ is larger at $E_{\text{c.m.}}$ below 1 MeV.  

In Fig.~4, we show $\sigma_{n,\gamma}$ as a function of $E_{\text{c.m.}}$ for 
the deformation parameter values of $\beta_2$ of 0.0, 0.2, 0.4, and 0.5, 
corresponding to a fixed $S_n$ of 0.35 MeV and the spectroscopic factor of 1. 
In this case sensitivity of the cross section to the deformation parameter 
is seen also for $E_{\text{c.m.}}$ below 1.0 MeV. We see that 
$\sigma_{n,\gamma}$ increases with increasing $\beta_2$, which reflects the 
trend seen in the $\beta_2$ dependence of the CD cross sections.
\begin{figure}[t]
\centering
\includegraphics[width=.45\textwidth]{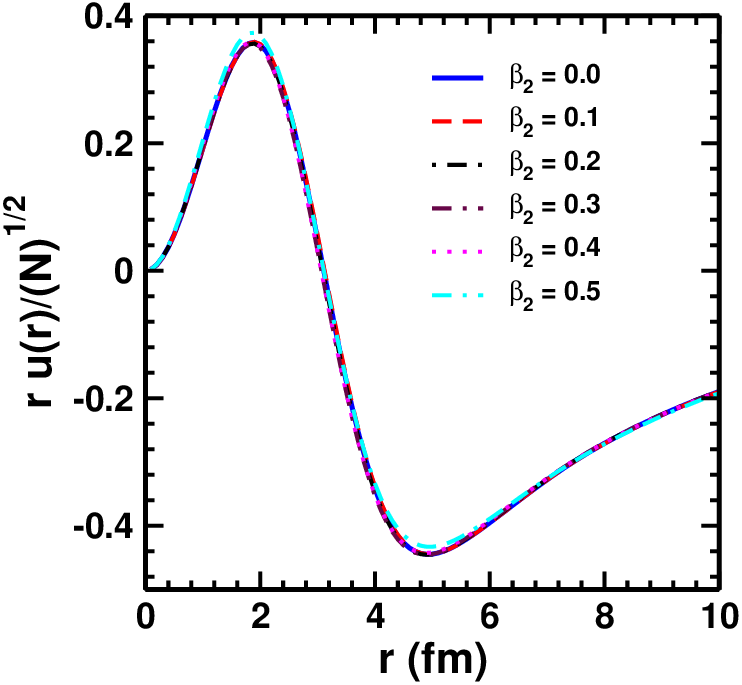}
\caption{Total wave function [$ru(r)$] for the $\ell$ = 1 and $j = 3/2$ state
including components with $\ell = 1, 3, 5$ and all the allowed $j$ values for 
different $\beta_2$ parameters. All the wave functions are normalized to unity.
}
\label{fig:Fig5}
\end{figure}

In our calculations the $\beta_2$ dependence of the CD cross sections and 
hence that of the capture cross section results primarily from the fact that
this parameter enters into the reaction amplitude explicitly through the 
potential $V_{bc}$ [see Eq.~(7)]. Had this not been the case, there would be no dependence of the cross sections on the $\beta_2$ as these peripheral 
reactions are governed mainly by the asymptotic normalization coefficient
(ANC), which is independent of the $\beta_2$, as is shown in Table I for the
$p_{3/2}$ state wave function.
\begin{table}[h]
\begin{center}
\caption {Asymptotic normalization constant (ANC) for the deformed 
$^{37}$Mg wave function for different values of the deformation parameter 
$\beta_2$. The deformed wave functions are obtained by solving the 
Sch\"odinger equation with potential given by Eq.~(7) with a $S_n$ value of 
0.35 MeV. Results are shown only for $\ell = 1$ and $j = 3/2$ state for 
each value of the deformation parameter.
}
\vspace{0.5cm}
\begin{tabular}{cccc}
\hline
\hline
$\beta_2$ & $\ell$ & $j$ & ANC \\
          &        &     & \footnotesize(fm$^{-1/2}$ )\\ 
\hline
0.0       & 1      & 3/2 & 0.383 \\
0.1       & 1      & 3/2 & 0.383 \\
0.2       & 1      & 3/2 & 0.380 \\ 
0.3       & 1      & 3/2 & 0.377 \\
0.4       & 1      & 3/2 & 0.374 \\
0.5       & 1      & 3/2 & 0.370 \\
\hline
\hline
\end{tabular}
\end{center}
\end{table}
\noindent 

Moreover, even the asymptotic part of the total wave function (which has 
contributions from components corresponding to $\ell = 1, 3$, and $5$ with all 
the allowed $j$ values), is unaffected by changes in $\beta_2$. This is 
illustrated in Fig.~5 where we plot the total wave function corresponding
to different values of $\beta_2$. The wave function in each case is normalized
to 1 to make the comparison easier and more meaningful. We see that the varying
of $\beta_2$ leaves the asymptotic part of the wave function completely unchanged,
which is in agreement with the results shown in Ref.~\cite{cap10}. This further 
strengthens the fact that the $\beta_2$ dependence of the CD cross section 
(and thus that of the capture cross section) is mainly due to the explicit 
presence of this parameter into the reaction amplitude. 
 
The results for the capture cross sections shown in Figs.~2 and 4 correspond 
to the CD cross sections obtained with the configuration of $^{36}$Mg($0^+$)
$\otimes$2$p_{3/2}n$ for a $C^2S$ value of 1. However, in Ref.~\cite{kob14} 
a $C^2S$ of 0.42${\pm 0.12}$ was deduced for this configuration from an 
analyses of the data on Coulomb breakup within a semiclassical theory of this 
reaction.  Even though this result is quite dependent on the theory of CD 
used in their analysis, had we used their value of $C^2S$ our results would 
have been proportionately lower.

Reaction rates ($R$) calculated from the capture cross sections are plotted 
in Figs.~6 and 7 as a function of $T_9$ (the temperature equivalent of relative 
energy in units of 10$^9$K, calculated from the relation $E_{\text{c.m.}} = k_BT$). We 
recall that the experimental cross sections for the Coulomb breakup reaction 
$^{37}$Mg + Pb $\rightarrow$ $^{36}$Mg + $n$ + Pb, involve uncertainties of about 
15-20$\%$~\cite{kob14}, which should also be there in the calculated Coulomb 
dissociation cross sections that are fitted to these data. Therefore, the 
$(n,\gamma)$ capture cross sections, and hence the rates of these reactions shown 
in Figs. 6 and 7, should also involve uncertainties of this order.

\begin{figure}[t]
\centering
\includegraphics[width=.45\textwidth]{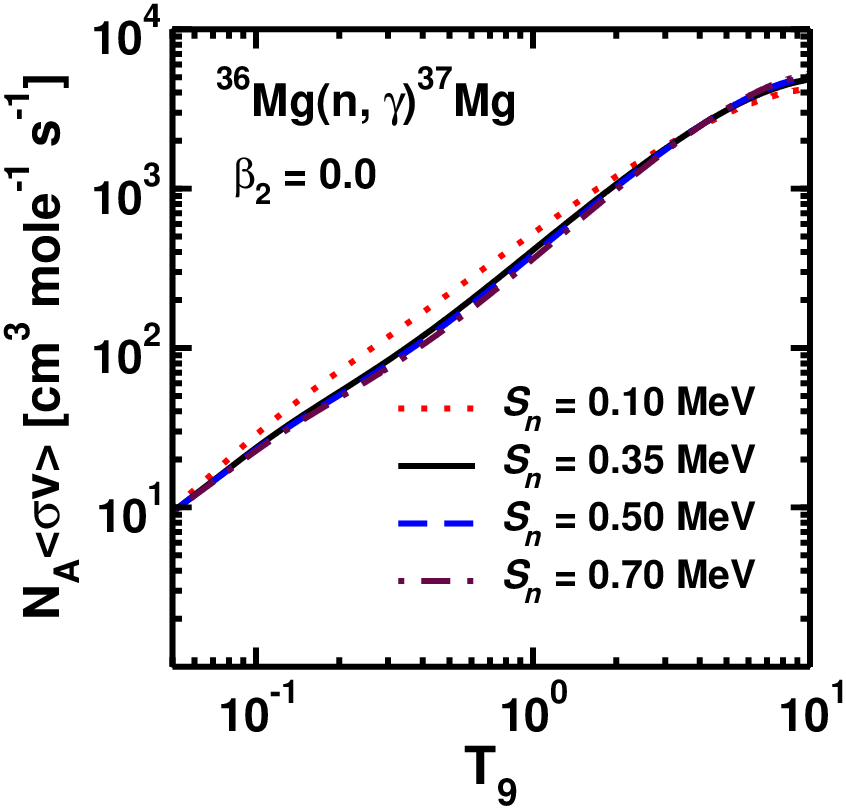}
\caption{Capture rates for the $^{36}$Mg($n, \gamma$)$^{37}$Mg reaction as a 
function of temperature in units of 10$^9$K ($T_9$) for different values of
$S_n$ for a fixed $\beta_2$ of 0.0.
}
\label{fig:Fig6}
\end{figure}
   
In Fig.~6 the reaction rate is shown for different values of $S_n$ for a fixed 
$\beta_2$ of 0.0, while in Fig.7 it is shown for various values of $\beta_2$ 
for a fixed $S_n$ of 0.35 MeV. The ground state configuration of $^{37}$Mg 
remains the same as that in Figs.~2 and 4. We note that the reaction rate changes  
from 10 cm$^3$ mole$^{-1}$ s$^{-1}$ to about 5000 cm$^3$ mole$^{-1}$ s$^{-1}$ as 
$T_9$ goes from 0.05 to 10. Its value around $T_9$ = 0.6 is approximately 200 
cm$^3$ mole$^{-1}$ s$^{-1}$. The $S_n$ and $\beta_2$ dependencies of the reaction 
rate reflect the trends seen in the dependencies of the capture cross section on 
these quantities, in Figs.~2 and 4. It may be noted that $T_9$ in the range of 
0.05 - 10 corresponds to $E_{\text{c.m.}}$ approximately in the range 4 keV to 1 
MeV.

As is evident from the integrand in Eq.~(1), for a fixed stellar temperature,
the maximum contribution to the reaction rate is strongly dependent on the
reaction cross section and in turn on the relative energy. This is substantiated
in Fig.~8, where we show the integrand of Eq.~(1) as a function of $E_{\text{c.m.}}$
for the reaction $^{36}$Mg($n, \gamma$)$^{37}$Mg at different values of $S_n$, 
but fixed $\beta_2$ and $T_9$ of 0.0 and 1, respectively. We see that maximum 
contribution to the rate of this reaction comes from $E_{c.m.}$ lying roughly 
between 0.2 - 0.3 MeV. At this low energy it is extremely difficult to measure 
reaction cross sections by direct methods. This is where the power of the 
CD method becomes evident as an indirect method in nuclear astrophysics. 
With the recent advances in experimental techniques it is possible to measure 
relative energy spectra at quite low relative energies in the CD experiments.

\begin{figure}[t]
\centering
\includegraphics[width=.45\textwidth]{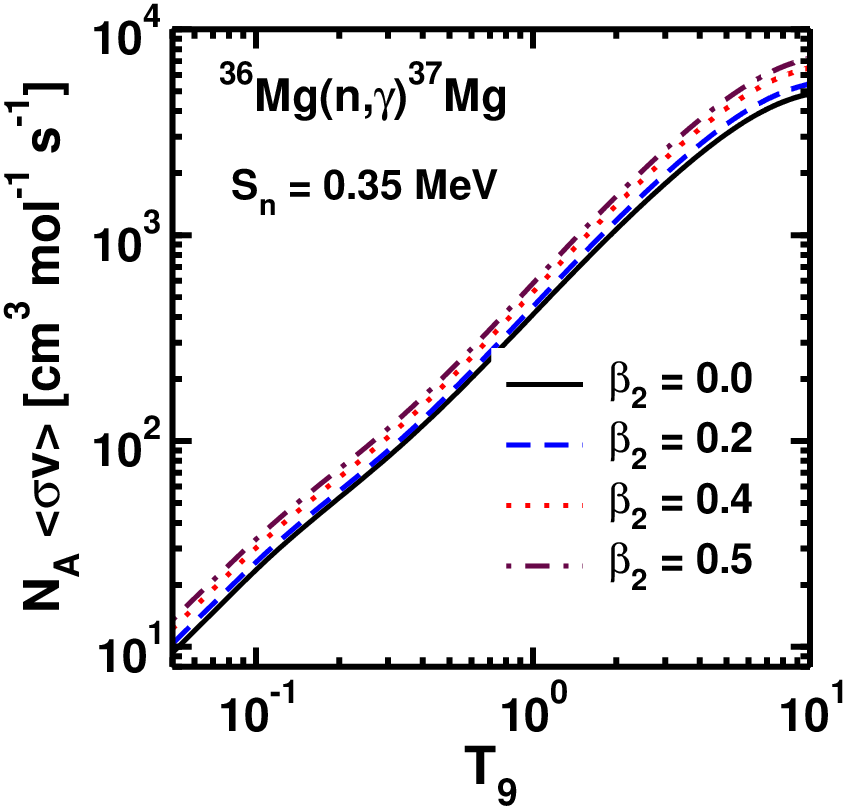}
\caption{Same as in Fig. 4  for different values of $\beta_2$ for a fixed 
$S_n$ of 0.35 MeV
}
\label{fig:Fig7}
\end{figure}

In Fig. 9, we show a comparison of the rates of $^{36}$Mg($n, \gamma$)$^{37}$Mg 
and $^{36}$Mg($\alpha, n$)$^{39}$Si reactions for the astrophysically relevant 
stellar temperature, $T_9$, in the range of 0.05 - 10. In calculations of the 
$(n,\gamma)$ reaction the value of $S_n$ is taken to be 0.35 MeV. Results are 
shown for $\beta_2$ of 0.0 and 0.5. The rates of the $(\alpha, n)$  reaction 
are calculated from the corresponding cross section given in Ref.~\cite{rau01} 
obtained from the NON-SMOKER code. We note that for $T_9 \geq 2$ the 
$^{36}$Mg($\alpha,n$)$^{39}$Si reaction is faster. Therefore, for these 
temperatures $\alpha$ capture reactions are more efficient and the formation 
of elements of higher charge number ($Z$) via the $\alpha$-induced processes 
is most important. However, at temperatures $T_9$ below 2, the ($\alpha,n$) 
reaction becomes progressively slower and the $(n,\gamma)$ reaction starts 
becoming more and more important. At these temperatures the classical 
$r$-process flow involving ($n,\gamma$) and ($\gamma,n$) reactions followed 
by $\beta$ decay is much more probable. In Fig.~9, we also note that the 
effect of projectile deformation is to increase the $(n,\gamma)$ rates 
slightly over the no deformation case, but this is insignificant as for 
as main conclusion of this figure is concerned.

\begin{figure}[t]
\centering
\includegraphics[width=.45\textwidth]{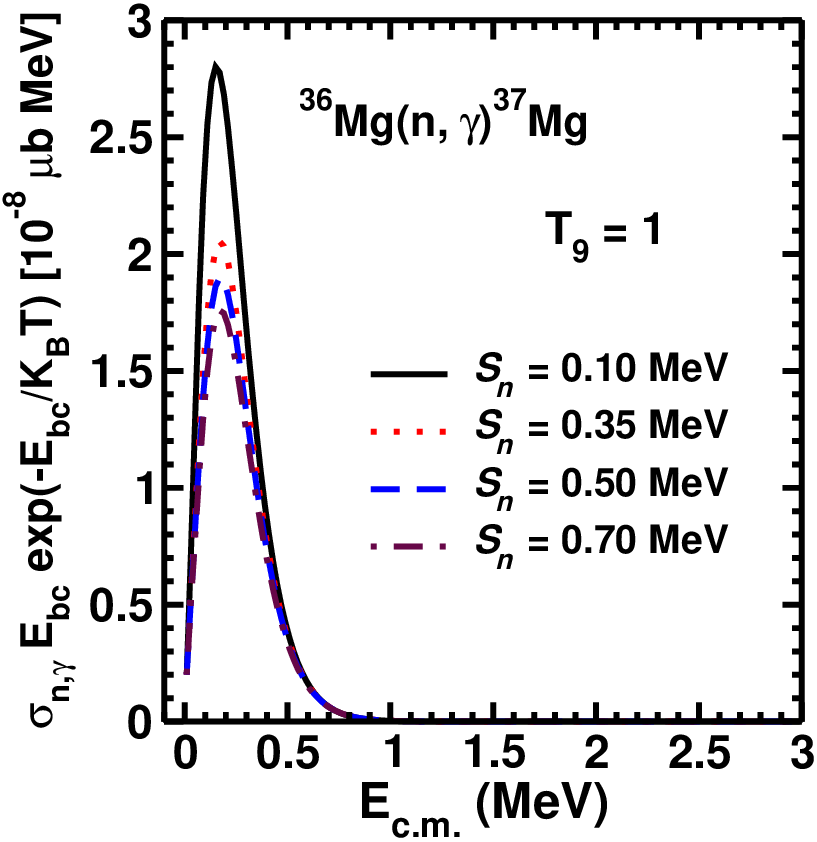}
\caption{Integrand of the reaction rate expression Eq. (\ref{Eq.1}) as a 
function of $E_{c.m.}$ for the $^{36}$Mg($n, \gamma$)$^{37}$Mg reaction at 
different values of $S_n$ for a fixed $\beta_2$ of 0.0. 
}
\label{fig:Fig8}
\end{figure}

Also shown in Fig.~9 are the rates of the $^{36}$Mg($n, \gamma$)$^{37}$Mg 
reaction obtained from the HF cross section reported in Ref.~\cite{rau01}.
We notice that CD $(n,\gamma)$ rates are significantly larger than those of 
the HF model. For $T_9 \leq 1$ the difference between CD and HF rates is 
quite drastic (several orders of magnitude). However, the difference between 
them becomes relatively lesser and lesser as $T_9$ increases beyond 1. The 
similar observation was also made in Ref.~\cite{tak05} for the case of the 
$^{18}$C($n, \gamma$)$^{19}$C reaction. This emphasizes the need for 
accurate determination of the rates of the $(n,\gamma)$ reaction on neutron 
rich light nuclei where the CD method can play a crucial role.

It is clear from Fig.~9 that around the equilibrium temperature, $T_9 
= 0.62$ where the main path of the reaction network runs through very 
neutron-rich nuclei, the $^{36}$Mg($n, \gamma$)$^{37}$Mg reaction is much 
faster than the $^{36}$Mg($\alpha,n$)$^{39}$Si reaction. Therefore, the 
$(n,\gamma)$ $\beta$-decay $r$-process is highly unlikely to be broken at 
the $^{36}$Mg isotope by the $\alpha$-process.

\begin{figure}[t]
\centering
\includegraphics[width=.45\textwidth]{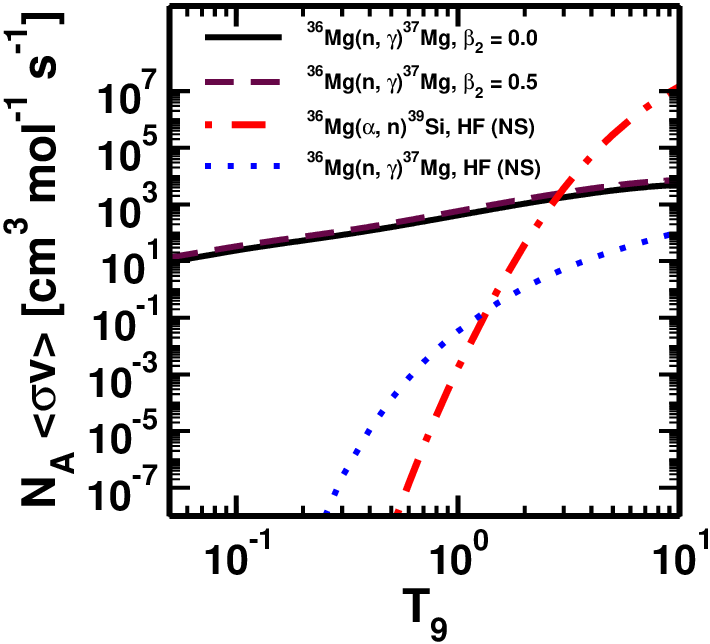}
\caption{Comparison of the rates of $^{36}$Mg($n, \gamma$)$^{37}$Mg 
reaction calculated by the CD method using $\beta_2$ parameters of 0 
(solid line) and 0.5 (dashed line) (the neutron separation energy                              
$S_n$ was 0.35 MeV in both cases), with those of the 
$^{36}$Mg($\alpha,n$)$^{39}$Si (dashed-dotted line) and 
$^{36}$Mg($n, \gamma$)$^{37}$Mg (dotted line) reactions obtained from 
the Hauser Feshbach (HF) cross sections adopted from the 
Ref.~\protect\cite{rau01} where they were obtained from the code 
NON-SMOKER(NS). The x-axis represents the temperature $T_9$.
}
\label{fig:Fig9}
\end{figure}

It is important to note that the while HF cross sections have contributions
only from the compound nuclear formations and decay mechanisms, the CD method 
produces only the direct capture component. In principle, both components 
coexist and need to be considered simultaneously when they are of the same 
order of magnitude.

\section{Summary and conclusions}

In summary, we calculate the rate of the $^{36}$Mg($n, \gamma$)$^{37}$Mg
reaction by studying the inverse photodissociation reaction in terms of the
Coulomb dissociation of $^{37}$Mg on a $^{208}$Pb target at the beam energy of 
244 MeV/nucleon using a theory formulated within the post-form finite range  
distorted-wave Born approximation that is extended to include the effects of 
the projectile deformation. This capture reaction is important in deciding if 
the $r$-process reaction flow will be sustained to Mg isotopes heavier than
$^{36}$Mg. If the rate of this reaction is smaller than that of the 
$^{36}$Mg($\alpha,n$)$^{39}$Si process, then the reaction flow will be broken
at this point, thereby reducing the production of Mg isotopes with mass numbers
larger than 36.

The advantage of our theoretical method is that it is free from the 
uncertainties associated with the multipole strength distributions of the 
projectile. In this approach measurements performed at beam energies in the 
range of few tens of MeV to few hundreds of MeV are used to extract cross 
sections of reactions at astrophysically relevant energies that usually lie 
in the range of few tens of keV to few hundreds of keV. Measurements performed at 
higher beam energies enhance the cross sections considerably. At higher 
energies the fragments in the final channel emerge with larger velocities, 
which facilitates their more accurate detection. By choosing adequate 
kinematical conditions of the coincidence measurements, it becomes possible 
to study the final state fragments at low relative energies, and to ensures 
that the target nucleus remains in the ground state during the reaction.
 
Our calculations suggest that the consideration of the deformation of the 
projectile nucleus in the Coulomb dissociation calculations, does not have any 
significant effect on the rate of the $^{36}$Mg($n, \gamma$)$^{37}$Mg reaction. 
Furthermore, the uncertainty in the value of one-neutron separation energy 
of the $^{37}$Mg nucleus also does not make any noticeable impact on the 
this rate. 
 
We find that for stellar temperatures $T_9$ above 2 the rates of the 
$^{36}$Mg($\alpha,n$)$^{39}$Si reaction are larger than those of the 
$^{36}$Mg($n, \gamma$)$^{37}$Mg reaction. This implies that at these 
temperatures the $\alpha$-capture reactions are more efficient than the 
neutron capture. Thus, $\alpha$-process operates at temperatures $T_9 \geq 2$. For lower temperatures ($T_9$ below 2) however, the $(\alpha,n)$  
reaction rates become progressively smaller than those of the $(n,\gamma)$ 
reaction. Eventually, the neutron capture becomes predominant and the 
classical $r$-process like flow, [$(n,\gamma)$ and $(\gamma,n)$ reactions 
followed by the $\beta$ decay], becomes the key process.

It may be remarked that the Hauser-Feshbach model, which is adopted by us 
to get the rates of the $^{36}$Mg($\alpha,n$)$^{39}$Si reaction, may not be  
a good approximation for the neutron rich nuclei. Nevertheless, we use these 
estimates because they are easy to obtain and their uncertainties are not 
larger than the differences seen between and $(n,\gamma)$ and $(\alpha,n)$ 
reaction rates in Fig.~9~\cite{per16}.

Near the saturation temperature $T_9 = 0.62$, the $(n,\gamma)$ reaction rate 
is several orders of magnitude larger than that of the 
$^{36}$Mg($\alpha,n$)$^{39}$Si reaction. Therefore, the $(n,\gamma)$ 
$\beta$-decay reaction flow is highly unlikely to be broken at the $^{36}$Mg
isotope and the reaction path of the $r$-process can go to Mg isotopes 
with mass numbers larger than 36 that are even closer to the corresponding 
neutron-drip line.

\section{acknowledgments}
 
This work was supported by the Science and Engineering Research Board (SERB),
Department of Science and Technology, Government of India under Grant Nos.
SR/S2/HEP-040/2012 and SB/S2/HEP-024/2013. One of us (Shubhchintak) is 
supported by the U.S. National Science Foundation (NSF) Grant No. PHY-1415656 
and the U.S. Department of Energy (DOE) Grant No. DE-FG02-08ER41533. We thank 
Peter Mohr for several useful correspondences.

\end{document}